\documentclass[epj,twocolumn]{webofc}
\usepackage[varg]{txfonts}   
\woctitle{TRANSVERSITY 2014}
\begin{document}
\title{Transverse Spin Effects in Future Drell-Yan Experiments}
%
%

\author{Jen-Chieh Peng\inst{1}\fnsep\thanks{\email{jcpeng@illinois.edu}}
}

\institute{Department of Physics, University of Illinois at 
Urbana-Champaign, Urbana, Illinois 61801, USA}

\abstract{%
We review the current status and future prospect for probing the transverse
momentum dependent (TMD) parton distributions using the Drell-Yan process.
We focus on the Boer-Mulders and Sivers functions, which are expected
to undergo a sign-change from semi-inclusive deep-inelastic scattering 
(SIDIS) to Drell-Yan process. The constraints of existing Drell-Yan and
SIDIS experiments on the signs of these functions are discussed.  
Future Drell-Yan measurements for the TMDs are also presented.
}
\maketitle
\section{Introduction}
\label{intro}
In this article, we review the current status and future prospect for
studying the transverse momentum dependent (TMD) parton
distributions in the nucleons using the Drell-Yan process. There has been
intense theoretical and experimental effort in recent decades to investigate
these novel parton distributions. There are many reasons for exploring
the transverse structures of the nucleons. For many years after the
discovery of scaling in DIS and partonic structures in the nucleons, only
the longitudinal momentum distributions of the partons were investigated.
The transverse degrees of freedom of the partons, accessed through the TMDs,
are expected to provide new insights on the nucleon structure.
The characteristics of the TMDs can also provide stringent tests for various
nucleon models. Moreover, the progress in lattice QCD also allows 
direct comparison between the lattice calculations and experiments. 

\begin{figure}
\centering
\includegraphics[width=8.0cm,clip]{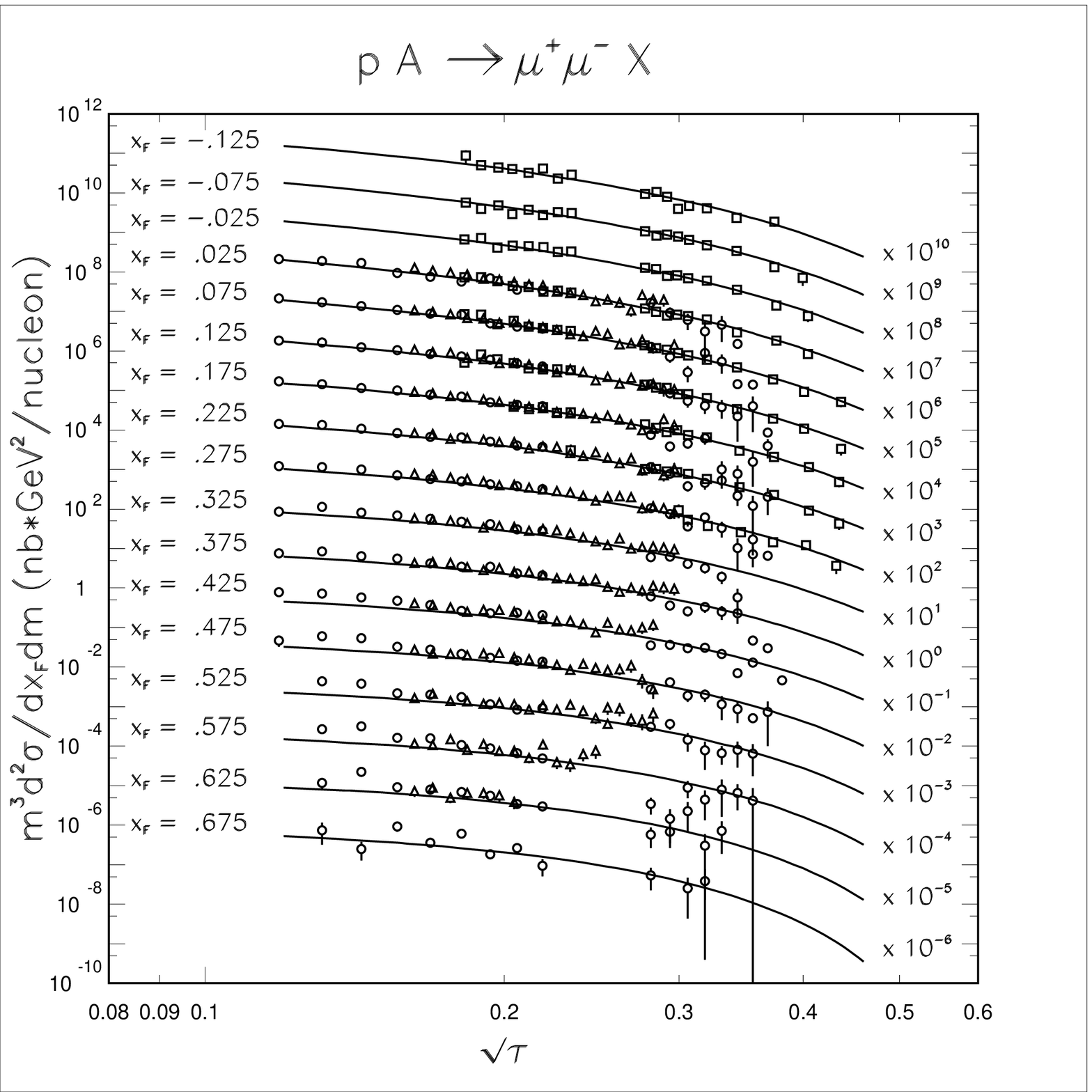}
\caption{Next-to-leading order Drell-Yan calculations compared to 
dimuon data
taken at various beam energies. From~\cite{plm}.}
\label{fig:DY}       
\end{figure}

The novel TMDs are accessible by experiments using either lepton or hadron
beams. The bulk of the experimental information on TMDs has been obtained
thus far from the semi-inclusive DIS (SIDIS) carried out at DESY (HERMES),
CERN (COMPASS), and the Jefferson Lab. A promising experimental tool
to access the TMDs is the Drell-Yan process~\cite{Drell70}. 
Proposed in 1970 to explain
the underlying mechanism for producing massive lepton pairs in
high-energy hadron-hadron collisions, the Drell-Yan process involves 
quark-antiquark annihilating into a virtual photon via electromagnetic
interaction. According to Yan~\cite{Yan98}, 
``The process has been so well understood
theoretically that it has become a powerful tool for precision measurements
and new physics". Indeed, the Drell-Yan process has provided unique information
on the antiquark distributions in nucleons and nuclei, as well as the
parton distributions of mesons and antiproton. Figure~\ref{fig:DY} shows
the proton-induced Drell-Yan cross sections measured at different beam
energies and NLO calculations~\cite{plm}. The Drell-Yan data 
taken at different
energies fall on smooth curves corresponding to the NLO calculations.
This is reminescent of the good agreement between DIS data and NLO
calculations. The well understood mechanism of the Drell-Yan process
provides a solid theoretical foundation for extracting novel parton 
distributions, like the TMDs, using this reaction.

To access the TMDs in the
Drell-Yan process would usually require transversely polarized beam
and/or target. Such transversely polarized Drell-Yan experiments have
never been performed yet, and are being pursued in many hadron
facilities. These polarized Drell-Yan experiments could indeed provide 
qualitatively new information on hadron physics.

In the following, we first discuss the unique features of the Drell-Yan
process in probing the TMDs. We then discuss the status and outstanding
issues in TMDs which could be addressed with future Drell-Yan experiments.
We conclude with a discussion on the status and prospect for measuring the
signs of the Sivers and Boer-Mulders functions in the Drell-Yan process.

\section{TMDs and the Drell-Yan Process}
\label{sec-1}
Comprehensive discussions on the theoretical and experimental aspects of
TMDs can be found in several review 
articles~\cite{Barone02,Barone10b,Aidala13}. 
In this paper, we focus on
three TMDs, the transversity distributions, the Boer-Mulders 
functions~\cite{Boer98}, and the Sivers functions~\cite{Sivers90}. 
There are three transverse quantities for nucleons
and quarks, namely, the nucleon's transverse spin ($\vec S^N_\perp$),
the quark's transverse spin ($\vec s^q_\perp$), and quark's transverse
momentum ($\vec k^q_\perp$). From these three quantities, one could form
three different correlations. The correlation between the quark's transverse
spin and nucleon's transverse spin is the transversity distribution. The
correlation between quark's transverse momentum and nucleon's transverse spin
leads to the Sivers function. Finally, the Boer-Mulders function corresponds
to the correlation between quark's transverse spin and its transverse
momentum. Although there are other TMDs, most of the recent progress
in TMD centered on these three distributions. In this paper, we focus
our discussion on these distributions.

The TMDs can be accessed via semi-inclusive DIS via measurements using
unpolarized beam and
target, polarized target, or polarized beam and polarized target. In leading
twist formulation, the different TMDs can be identified through their
distinct azimuthal angular distributions. For example, the Boer-Mulders
function would give a $\cos 2 \phi_h$ azimuthal distribution in unpolarized
SIDIS, where $\phi_h$ is the hadron azimuthal angle relative to the lepton
scattering plane. The transversity and Sivers function would yield a
$\sin (\phi_h + \phi_s)$ and $\sin (\phi_h - \phi_s)$ angular dependence,
respectively, in SIDIS with a transversely polarized target ($\phi_s$ refers
to the azimuthal angle of target spin direction relative to the lepton
scattering plane). An impressive progress has been made in the last
decade to extract the TMDs in several extensive SIDIS experimental
programs, reviewed in Ref.~\cite{Barone10b,Aidala13}.

TMDs can also be accessed via Drell-Yan experiments in unpolarized
hadron-hadron collision, singly polarized hadron-hadron collision,
or doubly polarized hadron-hadron collision. The earliest discussion
of probing TMDs via the Drell-Yan process was by Ralston and 
Soper~\cite{Ralston79},
who pointed out that the chiral-odd transversity distribution can be 
measured in doubly
transversely polarized p-p Drell-Yan process. Detailed discussions on
how various TMDs can be measured in polarized and unpolarized Drell-Yan,
including the W/Z boson production which can be considered as generalized 
Drell-Yan where the photon is replace by W/Z, have been discussed in the
literature~\cite{Boer99,Arnold09}. Following the demonstration that
TMDs can be measured via the SIDIS experiments, it is important to 
address the question "what can the Drell-Yan experiments offer in
probing and understanding the TMDs?". It turns out that the Drell-Yan
experiments can provide many unique information on TMDs complementary
to what the SIDIS can offer. 

First, the Drell-Yan cross section involves the convolution of two
parton distributions (one or both of which can be a TMD), while the
SIDIS cross section is a convolution of a parton distribution and
a fragmentation function. The absence of the fragmentation process
in the Drell-Yan implies that no information on the often poorly
known fragmentation functions is required to extract the TMDs.
This important feature of the Drell-Yan process allows a
truly independent measurement of TMDs and provide a very significant
check of the results obtained from SIDIS.
Second, the TMDs of mesons and antiprotons can not be accessed by
SIDIS since they are not available as targets. Fortunately,
they can be accessed via the Drell-Yan process since meson and
antiproton beams are available. This also applies to hyperons,
although no hyperon beams are currently available anywhere.
Third, as demonstrated in existing Drell-Yan experiments, the proton-induced
Drell-Yan process is sensitive to the antiquark distributions in the
nucleons. While the effects of the nucleon's antiquark TMDs are usually
overshadowed by those of the valence quark's TMDs in SIDIS, the antiquark's
TMDs of the nucleons could be sensitively probed in the Drell-Yan process.
Finally, the celebrated prediction that the time-reversal-odd
TMDs, namely, the Sivers and the Boer-Mulders functions, must undergo
a sign reversal between the space-like SIDIS and the time-like Drell-Yan
reactions awaits experimental confirmation. This has become a major physics
goal of several proposed polarized Drell-Yan experiments.

\section{Status and Outstanding Issues in TMDs}

\subsection{Transversity}

The transversity distribution, described in the quark-parton model as the
net transverse polarization of quarks in a transversely polarized nucleon,
can be measured in SIDIS using transversely polarized targets. The azimuthal
angular modulation in $\sin(\phi_h + \phi_s)$ is proportional to $h_1
\otimes H^\perp_1$, which is the convolution of transversity and the
Collins fragmentation function. An important ingredient for extracting the
transversity distribution is the measurement of a sizable
Collins fragmentation function at Belle~\cite{Seidl08}. Recent SIDIS
experiments using transeversely polarized targets have been carried out
at HERMES~\cite{Airapetian05,Airapetian10}, COMPASS~\cite{Ageev07,Adolph12},
and JLab~\cite{Qian11,Zhao14}. 

\begin{figure}
\centering
\includegraphics[width=7.5cm,clip]{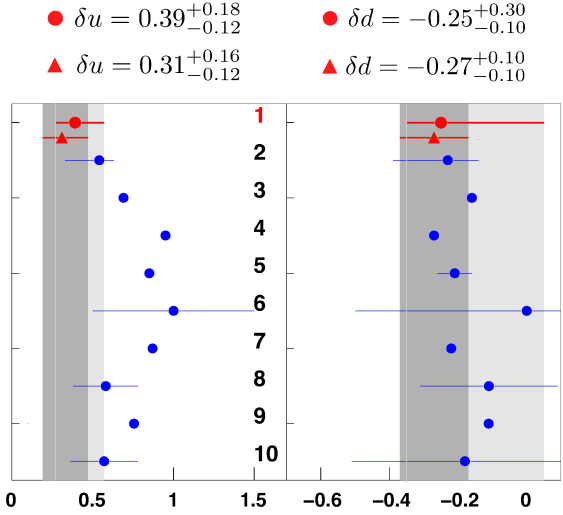}
\caption{Tensor charges $\delta u$ and $\delta d$ obtained from
the fits to the semi-inclusive DIS data at $Q^2 =2.4$ GeV$^2$ in
comparison with various calculations. From~\cite{Anselmino13}.}
\label{fig_trans}       
\end{figure}

A recent global analysis of the SIDIS and 
Collins fragmentation function data has led to the extraction of the
transversity distributions of $u$ and $d$ quarks~\cite{Anselmino13}.
This global analysis assumes the sea-quark transversity to be zero, and
it shows that $u (d)$ valence quark transversity distributions
are positive (negative),
in agreement with their corresponding helicity distributions. 
However, the magnitudes of
the transversity distributions are significantly smaller than the
corresponding helicity distributions. The tensor charges $\delta u$ and
$\delta d$, defined as $\delta q \equiv \int^1_0 [\delta q(x) - \delta
\bar q(x)]dx$, have also been determined from this analysis, as shown in
Fig.~\ref{fig_trans}. The central values of the tensor charges are
$\delta u = 0.31$ and $\delta d = -0.27$, significantly smaller 
in absolute magnitude than the axial charges, $\Delta u = 0.787$
and $\Delta d = -0.319$. As shown in Fig.~\ref{fig_trans}, the extracted
value of $\delta u$ is significantly smaller than the predictions of
various models and lattice calculations~\cite{Anselmino13}. 

Since sea quark transversity is neglected in the recent extraction
of transversity distributions~\cite{Anselmino13}, it could introduce 
systematic uncertainties
in the determination of the tensor charges $\delta u$ and $\delta d$.
It is interesting to examine the theoretical predictions on the
sign and magnitude of the sea-quark transversity. In the non-relativistic
limit, one expects the same sea-quark transversity and helicity 
distributions. Model predictions, however, do not necessarily follow this
non-relativistic expectation. In the large $N_c$ limit~\cite{Schweitzer01}, 
the chiral-quark soliton model predicts~\cite{Wakamatsu07} 
that $\delta \bar u(x) - \delta \bar d(x) < 0$,
which is opposite to the case of helicity distribution, 
$\Delta \bar u(x) - \Delta \bar d(x) > 0$. 
A qualitatively different prediction of 
$\delta \bar u(x) - \delta \bar d(x) > 0$
is obtained in a statistical model~\cite{Bourrely09}. Very recently,
an interesting new approach to calculate the $x$ dependence of the
isovector $\delta \bar u(x) - \delta \bar d(x)$ in lattice QCD indicated
large negative values~\cite{Lin13}. It would certainly be 
very interesting to measure
$\delta \bar u(x)$ and $\delta \bar d(x)$ to test the predictions of these
models.

It is clear that remarkable progress has been made in measuring the
illusive chiral-odd transversity distributions via SIDIS. Some remaining
challenges to be addressed by future experiments include

\begin{itemize}

\item Independent measurements of the transversity distributions. There
exist other reactions involving two chiral-odd functions which are sensitive 
to the transversity distributions. First results~\cite{Airapetian08,Adolph12a}
on dihadron SIDIS using
the corresponding interference fragmentation functions determined at 
Belle~\cite{Vossen11} are becoming available. The extracted transversity
distributions~\cite{Bacchetta11} are consistent with those extracted in
single-hadron SIDIS~\cite{Anselmino13}. Additional dihadron SIDIS data 
are expected using the SoLID detector at the 12 GeV JLab. The proposal
to measure Drell-Yan with polarized antiproton beam colliding with 
polarized proton at FAIR~\cite{Barone05} would clearly provide 
a much anticipated
measurement of transversity entirely free of the uncertainty of fragmentation
functions.

\item Magnitude and sign of the sea-quark transversity distribution.
As discussed above, some theoretical models and the recent lattice calculation
suggest sizable sea-quark transversity. A recent 
SIDIS measurement~\cite{Zhao14} at JLab
shows a large negative $\sin(\phi_h+\phi_s)$ amplitude for $K^-$
production on a transversely polarized $^3$He target, suggesting a
large sea-quark transversity. Future high-statistics SIDIS experiments
at the 12 GeV JLab upgrade, as well as doubly polarized Drell-Yan
experiments in $pp$ collision, uniquely sensitive to the antiquark
transversity distribution in the proton, are required to pin down
the role of sea quarks in transversity distributions.

\item $Q^2$ evolution of the transversity distributions. The absence of
the gluon transversity distribution implies a slower $Q^2$-evolution
for the quark transversity distributions. This expectation remains to
be tested. The challenge in the SIDIS, however, is to disentangle the
$Q^2$-evolution effect of the Collins fragmentation function. A more
definitive measurement would require the Drell-Yan process, which
only involves the convolution of two transversity distributions.

\end{itemize}

\subsection{Sivers Functions}

It was suggested by Sivers that single-spin asymmetries in various processes
can originate from the correlations between the transverse momentum of the
quark and proton's transverse spin, called the Sivers function~\cite{Sivers90}. 
As a time-reversal odd
object, the Sivers function requires initial- or final-state interactions
via a soft gluon. Such interactions are incorporated in a natural fashion
by the gauge link needed for a gauge-invariant definition of the 
TMD~\cite{Brodsky02,Collins02,Ji02}. 
The Sivers function is related to the forward scattering amplitude in
which the helicity of the nucleon is flipped. This helicity flip must involve
the orbital angular momentum of unpolarized quark. In this respect, 
the Sivers 
function is connected to the angular momentum of the quark. An unambiguos
measurement of Sivers function is valuable for understanding the nature of
the TMDs and the spin content of the nucleons.

Measurements of the $\sin(\phi_h - \phi_s)$ Sivers moment in polarized
SIDIS with transversely polarized targets have been performed 
by HERMES~\cite{Airapetian09}, COMPASS~\cite{Alekseev09,Alekseev10}, and
the JLab Hall-A~\cite{Qian11} collaborations. The quark and antiquark
Sivers functions were extracted in recent global fits to the  
data~\cite{Anselmino09,Sun13}. The analysis confirms the theoretical
expectations for the signs of the $u$ and $d$ Sivers functions.
Non-zero $\bar d$ sea-quark Sivers function is also obtained in order to
explain the large Sivers moment observed for $K^+$. In the meson-cloud
model, the pseudo-scalar meson is in a $p$ state. Since meson contains
valence antiquarks, it is reasonable to expect that antiquarks carry
non-zero orbital angular momentum. Using chiral-quark solition model,
it was shown that $\bar u$ and $\bar d$ have significant contributions
to the orbital angular momentum component of the proton's 
spin~\cite{Wakamatsu10}. Moreover, a recent lattice QCD 
calculation~\cite{Deka13} found a significant
fraction of proton's spin comes from the orbital angular momentum of
$\bar u$ and $\bar d$ quarks. These results suggest that the Sivers
functions for sea quarks can be sizable and measurable. Future
polarized SIDIS experiments at JLab and EIC are expected to provide
definitive measurements on the characteristics of the sea-quark
Sivers functions.

Unlike collinear parton distribution functions (PDFs), the TMDs are not
necessarily universal and could depend on the process they are
extracted from. Fortunately, the process-dependence of TMDs is
limited to a possible sign-change, due to the invariance of QCD
under parity and time-reversal~\cite{Collins02}. The TMD,
$f_{q/p\uparrow}(x,k_\perp,\vec S)$, for finding
an unpolarized quark inside a transversely polarized proton extracted
from SIDIS and Drell-Yan proces satisfies~\cite{Collins02,Collins04}
\begin{equation}
f^{SIDIS}_{q/p\uparrow}(x,k_\perp,\vec S) = 
f^{DY}_{q/p\uparrow}(x,k_\perp,-\vec S).  
\label{eq1}
\end{equation}
The Sivers function, $f^\perp_{1T}(x,k_\perp)$, is proportional 
to $[f_{q/p\uparrow}(x,k_\perp,\vec S)
-f_{q/p\uparrow}(x,k_\perp,-\vec S)]/2$. Therefore, Eq.~\ref{eq1}
implies that the Sivers function changes sign from SIDIS to
Drell-Yan
\begin{equation}
f^{\perp~SIDIS}_{1T}(x,k_\perp) = - f^{\perp~DY}_{1T}(x,k_\perp)
\label{eq2}
\end{equation}
This sign-change prediction, which is a consequence of 
factorization of TMD in QCD, has generated
tremendous interest to test it experimentally.
Since the signs of valence quark Sivers
function have already been determined from SIDIS, the test only
requires measurements of valence-quark Sivers functions in
singly polarized Drell-Yan experiments utilizing transversely polarized
proton beam or target. As discussed by Chiosso in this 
Workshop~\cite{Chiosso14}, a
dedicated Drell-Yan experiment at COMPASS using 190 GeV pion beam on
transversely polarized target is scheduled to take data 
in 2015~\cite{Denisov09}.
Teryaev also presented the plan to measure polarized Drell-Yan 
at NICA~\cite{Teryaev14}.
Other polarized Drell-Yan experiments, proposed
at existing or future hadron facilities, include RHIC at BNL~\cite{Bunce00},
PAX~\cite{Barone05} and PANDA~\cite{Brinkmann07} at FAIR, J-PARC~\cite{Peng00}, 
E1027~\cite{P1027} and E1039~\cite{P1039} at Fermilab.
While it is highly anticipated that the first result on this test 
would be obtained at COMPASS~\cite{Chiosso14}, it is essential 
to perform experiments
using different hadron beams (pion, proton, antiproton) covering 
different kinematic regions in $x$ and $Q^2$.

\subsection{Boer-Mulders Functions}

The Boer-Mulders function~\cite{Boer98},
$h^\perp_1(x,k_\perp)$, is another example of a time-reversal odd TMD. It 
represents the correlation
between $\vec k_\perp$ and the quark transverse spin, $\vec s_\perp$, 
in an unpolarized
nucleon. Similar to 
the Sivers function, the Boer-Mulders function 
also owes its existence to the presence of initial/final state
interactions. 

The flavor and $x$ dependencies of the  
Boer-Mulders functions have been calculated in various models.
In the quark-diquark model 
taking into account both the scalar and axial-vector  
diquark configurations, significant
differences in the flavor dependence are found between the Sivers and 
Boer-Mulders functions~\cite{Gamberg08}. In particular, while 
the Sivers function is negative for the $u$ and positive for the 
$d$ valence quarks, the $u$ and $d$ valence quark Boer-Mulders
functions are predicted to be both negative. Other model calculations
using the large-$N_c$ model~\cite{Pobylitsa03}, the MIT bag 
model~\cite{Yuan03}, the relativistic constituent
quark model~\cite{Pasquini07}, as well as lattice QCD~\cite{Gockeler07},
all predict the $u$ and $d$ valence Boer-Mulders
functions to be negative. 

While the Sivers functions do not exist
for spin-zero hadrons, the Boer-Mulders functions can exist for pions.
Calculations for pion's valence-quark Boer-Mulders functions
using the quark-spectator-antiquark model
predict a negative sign~\cite{Lu05},
just like the $u$ and $d$ Boer-Mulders functions for nucleons. 
Using the bag model, the valence 
Boer-Mulders functions for nucleons and mesons 
were predicted~\cite{Burkardt08} to have same signs with similar 
magnitude. This prediction of a universal
behavior of the valence-quark Boer-Mulders functions for pions and nucleons 
also awaits experimental confirmation.

For nucleon's antiquark Boer-Mulders functions there
exists one calculation using the meson cloud model~\cite{Lu06}.
Since the meson cloud is an important source for sea quarks in the
nucleons, as evidenced by the large $\bar d / \bar u$ flavor
asymmetry observed in DIS and Drell-Yan experiments (for a
recent review see~\cite{Chang14}), the valence Boer-Mulders functions
in the pion cloud can contribute to nucleon's
antiquark Boer-Mulders functions~\cite{Lu06}. An interesting prediction
is that nucleon's antiquark Boer-Mulders functions would have 
negative signs, same as the
valence Boer-Mulders functions for pion.

For a TMD with a tensor spin projection, 
$f_{h_{1q}/p\uparrow}(x,k_\perp,\vec S)$,
Eq.~\ref{eq1}, which relates the TMDs measured in SIDIS and Drell-Yan,
becomes~\cite{Collins04,Peng14}
\begin{equation}
f^{SIDIS}_{h_{1q}/p\uparrow}(x,k_\perp,\vec S) =
- f^{DY}_{h_{1q}/p\uparrow}(x,k_\perp,-\vec S).
\label{eq3}
\end{equation}
The minus sign in Eq.~\ref{eq3} relative to Eq.~\ref{eq1} is due to the
replacement of the vector spin projection $\gamma^+$ in the TMD 
$f_{q/p\uparrow}(x,k_\perp,\vec S)$ by the tensor spin projection 
$\sigma^{+\perp}$ in the TMD $f_{h_{1q}/p\uparrow}(x,k_\perp,\vec S)$.
Since the Boer-Mulders function is proportional to the sum
$[f_{h_{1q}/p\uparrow}(x,k_\perp,\vec S) + 
f_{h_{1q}/p\uparrow}(x,k_\perp,-\vec S)]/2$,
it follows from Eq.~\ref{eq3} that the Boer-Mulders function measured in
Drell-Yan will have a sign opposite to that measured in SIDIS, namely,
\begin{equation}
h^{\perp~SIDIS}_{1q}(x,k_\perp) = - h^{\perp~DY}_{1q}(x,k_\perp).
\label{eq4}
\end{equation}

\begin{table}   
\centering
\caption {Theoretical predictions for the signs of various 
Boer-Mulders functions
for nucleons ($N$) and pions ($\pi$) in SIDIS and Drell-Yan.
$V_\pi$ signifies the valence
quarks in the pions. The paranthesis for $V_\pi$ in SIDIS indicates
that it can not be measured in practice.}
\label{tab:sign}
\begin{center}
\begin{tabular}{|c|c|c|c|c|c|}
\hline
\hline
 & ~$u_N$~ & ~$d_N$~ & ~$V_\pi$~ & ~$\bar u_N$~ & ~$\bar d_N$~ \\
\hline
\hline
SIDIS & $-$ & $-$ & ($-$) & $-$ & $-$ \\
\hline
~~Drell-Yan~~ & + & + & + & + & + \\
\hline
\hline
\end{tabular}
\end{center}
\end{table}

Table~\ref{tab:sign} summarizes the theoretical expectations relevant to
the signs of the Boer-Mulders functions. First, the $u$ and $d$
Boer-Mulders functions of the nucleons have negative signs. Second, the
valence Boer-Mulders functions in the pions have the same signs as those of the
nucleons, namely, negative. Third, the antiquark Boer-Mulders
functions in the nucleons are also negative. Finally, the signs
of these Boer-Mulders functions will reverse and become positive
for the Drell-Yan process. We discuss next the experimental
status on extracting the Boer-Mulders functions from the SIDIS and
Drell-Yan experiments, and we compare the predictions listed in
Table~\ref{tab:sign} with data. We also
identify future experiments which can further test these predictions.

\subsubsection{Boer-Mulders functions from SIDIS}

The Boer-Mulders functions can be extracted from the hadron's azimuthal angular
distribution in unpolarized SIDIS.
The $\cos 2\phi$ term is proportional to the product
of the Boer-Mulders function $h^\perp_1$ and the Collins fragmentation
function $H^\perp_1$ at leading twist, where $\phi$ 
refers to the angle between the hadron plane and the lepton plane. 
The $\langle \cos 2 \phi \rangle$ moments have been measured by the
HERMES~\cite{Giordano09,Hermes13} and COMPASS~\cite{Kafer08,Bressan09}
collaborations. An analysis of the pion SIDIS
data, taking into account the higher-twist 
Cahn effect~\cite{Cahn78}, has been performed~\cite{Barone10a}.
The functional form for the Boer-Mulders function was assumed as
\begin{equation}
h_1^{\perp q} (x,k_\perp^2) = \lambda_q f_{1T}^{\perp q} (x, k_\perp^2),
\label{eq5}
\end{equation}
\noindent where $h_1^{\perp q}$ and
$f_{1T}^{\perp q}$ are the Boer-Mulders and Sivers functions, respectively,
for quark $q$.
The best-fit values are found to be $\lambda_u = 2.0 \pm 0.1$
and $\lambda_d = -1.111 \pm 0.001$. 
Since the Sivers function for $u (d)$
is negative (positive), these values show that $h_1^{\perp u}$ and
$h_1^{\perp d}$ are both negative, in agreement with the theoretical
expectation listed in Table~\ref{tab:sign}. It must be 
cautioned that the extracted signs of the 
Boer-Mulders functions depend on the signs of the Collins functions adopted in
the analysis. Although the signs chosen for the Collins functions
are based on plausible arguments, some uncertainties do exist
in the determination of the signs of the Boer-Mulders functions from
this analysis.

New results on the 
azimuthal $\cos 2 \phi$ modulations for $\pi^{\pm}$, $K^{\pm}$,
and unidentified hadrons in unpolarized $e+p$ and $e+d$ SIDIS were
recently reported by HERMES~\cite{Hermes13}.
These new HERMES data are expected to provide a more precise extraction of the
valence Boer-Mulders functions. Moreover, these data are sensitive to
the sea-quark Boer-Mulders functions. In particular, the $\cos 2 \phi$
moments for $K^-$ are observed to be large
and negative~\cite{Hermes13}. 
Since the valence quark content of $K^-$,
$s \bar u$, is different from those of the nucleons, the large
$K^-$ $\cos 2 \phi$ moments suggest sizable sea-quark 
Boer-Mulders functions.
An extension of the global fit in Ref.~\cite{Barone10a} to include the
new HERMES data could lead to a determination of the sign and magnitude
of sea-quark Boer-Mulders functions in SIDIS.

\subsubsection{Boer-Mulders functions from Drell-Yan}

The Boer-Mulders functions can be extracted~\cite{Boer99}
from the azimuthal angular distributions in unpolarized Drell-Yan
process. The expression for the Drell-Yan
angular distribution is
\begin{equation}
\frac {d\sigma} {d\Omega} \propto 1+\lambda \cos^2\theta +\mu \sin2\theta
\cos \phi + \frac {\nu}{2} \sin^2\theta \cos 2\phi,
\label{eq:DY_ang}
\end{equation}
\noindent where $\theta$ and $\phi$ correspond to the polar and azimuthal 
decay angles
of the $l^+$ in the dilepton rest frame. Boer showed that the $\cos 2\phi$
term is proportional to the convolution of the quark and antiquark
Boer-Mulders functions in the beam and target hadrons~\cite{Boer99}.
The $\cos 2\phi$ angular dependence can be qualitatively understood by 
noting that the Drell-Yan cross
section depends on the transverse spins of the annihilating quark and
antiquark. A correlation between the transverse spin and
the transverse momentum of the quark, as represented by the Boer-Mulders
function, would lead to a correlation between the transverse spin and
transverse momentum of the dimuons, resulting in the $\cos 2\phi$ 
dependence.

\begin{figure}[htb]
\begin{center}
\includegraphics[width=0.45\textwidth]{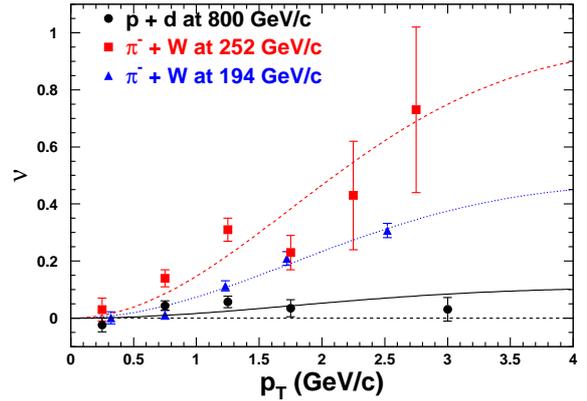}
\end{center}
\caption{The $\cos 2 \phi$ amplitude $\nu$ for pion- and
proton-induced Drell-Yan measurements. Curves are fits to
the data using an empirical parametrization discussed in 
Ref.~\cite{Zhu07}.} 
\label{e866_bm}
\end{figure}

Pronounced $\cos 2 \phi$ dependence,
observed in the NA10~\cite{Falciano86} and E615~\cite{Conway89}
pion-induced Drell-Yan experiments, was attributed to the
Boer-Mulders function. The $\cos 2 \phi$
dependence of proton-induced Drell-Yan process was also 
measured for the 
$p+d$ and $p+p$ interactions
at 800 GeV/c~\cite{Zhu07,Zhu09}. Unlike the pion-induced Drell-Yan,
significantly smaller (but non-zero) cos$2\phi$ azimuthal angular dependence
was observed in the $p+d$ and $p+p$ reactions, as shown in
Fig.~\ref{e866_bm}. While the pion-induced 
Drell-Yan cross section
is dominated by a valence antiquark in the pion
annihilating a valence quark in the nucleon, the
proton-induced Drell-Yan process must involve 
a sea antiquark in the nucleon. Therefore, the
$p+d$ and $p+p$ results suggest~\cite{Zhu07,Zhu09} that 
proton's Boer-Mulders 
functions for
sea quarks are smaller than those for valence quarks.

\begin{figure}[tbp]
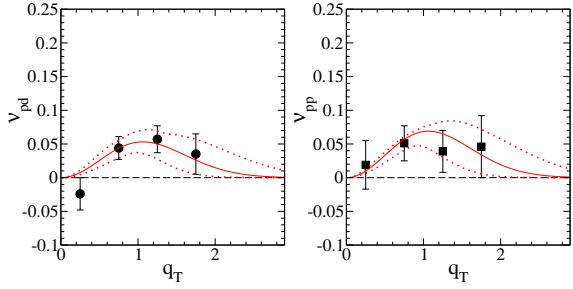

\begin{center}
\includegraphics*[width=0.45\linewidth]{pd_cos2phi.eps}
\includegraphics*[width=0.45\linewidth]{pp_cos2phi.eps}
\end{center}
\caption{Fits to the $\cos 2 \phi$
amplitude $\nu$ for the
$p+d$ and $p+p$ Drell-Yan data~\cite{Lu10}. The dotted lines
indicate the uncertainty of the fitted parameters. $q_T$
in unit of GeV refers to the transverse momentum of the dimuon pair.}
\label{e866_bm_fit}
\end{figure}

Several authors have extracted Boer-Mulders functions from 
the $p+p$ and $p+d$ Drell-Yan data~\cite{Zhang08,Lu10,Barone10}. 
The results obtained by Lu and Schmidt~\cite{Lu10} are
shown in Figure~\ref{e866_bm_fit}. Although the statistical
precision of the data does not yet allow an accurate extraction of 
the Boer-Mulders functions, the analysis shows that
both the $\bar u$ and $\bar d$ Boer-Mulders functions can be
extracted from the Drell-Yan. Ongoing unpolarized
proton- and pion-induced Drell-Yan experiments at Fermilab and
COMPASS are expected to provide new information on the 
Boer-Mulders functions.

\subsubsection{prospect for tesing the sign-reversal of Boer-Mulders functions}

While the prospect for testing the sign-reversal of
the Sivers functions has been
discussed extensively in the literature,
much less attention has been devoted to examining the possibility for
checking the sign-reversal for the Boer-Mulders functions.
This is possibly due to the fact that the Boer-Mulders functions are
just beginning to be extracted from SIDIS data.
On the other hand, while the Sivers functions have only been extracted from
the SIDIS data, information on the Boer-Mulders
functions has been obtained from both the SIDIS and the 
Drell-Yan experiments. In fact,
Boer-Mulders functions are the only TMD functions
ever measured in the Drell-Yan experiments so far.
It is natural to ask whether
these existing data  have already tested the prediction that
the Boer-Mulders function changes sign from SIDIS to Drell-Yan.

As shown in Table 1, theoretical models predict
that the valence quark Boer-Mulders functions have negative signs
in SIDIS. This prediction is consistent with the analysis of the existing
SIDIS data on the $\cos 2\phi$ angular dependence, showing that 
both $h^\perp_{1u}$ and $h^\perp_{1d}$ in SIDIS are 
negative~\cite{Barone10}. Unfortunately, SIDIS data
do not yet allow the extraction of sea-quark Boer-Mulders functions,
whose effects are overshadowed by the valence
quarks. Nevertheless, the $\cos 2\phi$ dependence in $p+p$ and
$p+d$ Drell-Yan is proportional to the convolution of the valence
and sea Boer-Mulders functions, namely,
$\nu \sim h^\perp_{1q} (x_1) h^\perp_{1\bar q} (x_2)$,
allowing a direct extraction of sea-quark Boer-Mulders function
in the Drell-Yan experiment. Assuming $u$-quark dominance,
the positive values of $\nu$ shown in Fig.~\ref{e866_bm_fit} 
already suggest that the Boer-Mulders functions for $u$ and $\bar u$
have the same sign in Drell-Yan. The
Boer-Mulders function for $u$ quark, determined in SIDIS
to have negative sign, is expected to become positive 
in Drell-Yan. It follows that both the $u$ and $\bar u$ 
Boer-Mulders functions in Drell-Yan should 
have positive signs. This is consistent with the
results obtained by Lu and Schmidt~\cite{Lu10} and
by Barone et al.~\cite{Barone10}. However, one can not
exclude the alternative possibility that both $u$ and $\bar u$ Boer-Mulders
functions in Drell-Yan have negative signs. 
At this moment,
one can only conclude that all existing SIDIS and Drell-Yan data are 
not in disagreement
with the predictions shown in Table 1.
Future SIDIS and Drell-Yan data are required to provide a definitive
test on the sign-change of the Boer-Mulders functions.

\subsection{Summary}

The massive Drell-Yan process in hadron-hadron collisions 
continues to be a powerful tool for probing the partonic structure of
hadrons and for understanding the dynamics of QCD in
perturbative and nonperturbative regimes. The Drell-Yan process is 
complementary to the DIS
in many aspects, and together, they provide interesting and
often surprising information on the 
quark and antiquark contents of the nucleons.  

The Drell-Yan process is also complementary to
SIDIS in extracting TMDs.  It is necessary to use both the 
Drell-Yan process and SIDIS to study TMDs, 
because of the non-universality of TMDs and the expected sign change.  
In this respect, the Sivers and Boer-Mulders functions are
the most interesting TMDs to explore. 
The Drell-Yan production of massive lepton pairs with polarized beam 
and/or target provide the immediate access to these two TMDs.  
With several ongoing and future experiments
at existing or future hadron facilities dedicated to measuring
singly or doubly polarized Drell-Yan, the full potential of the 
Drell-Yan process in studying TMDs is just beginning to be explored.

\end{document}